\documentclass[conference]{IEEEtran}
\usepackage{cite}
\usepackage{graphicx}
\usepackage{pgfplots}
\usepgfplotslibrary{groupplots}
\pgfplotsset{width=\linewidth,compat=1.9}
\usepackage{url}
\usepackage{tabularx, booktabs}
\usepackage{listings}
\usepackage{xcolor}
\usepackage[utf8]{inputenc}
\usepackage{makecell} 						
\usepackage[acronym]{glossaries}
\usepackage[nointegrals]{wasysym}			
\usepackage[shortlabels]{enumitem}			
\usepackage{csquotes}

\definecolor{light-gray}{gray}{0.95}
\lstdefinestyle{mystyle}{
    backgroundcolor=\color{light-gray},  
    basicstyle=\ttfamily\small
}

\newacronym{AAS}{AAS}{Asset Administration Shell}
\newacronym[shortplural={CSS}, longplural={Capabilities, Skills and Services}]{CSS}{CSS}{Capability, Skill and Service}
\newacronym{LLM}{LLM}{Large Language Model}
\newacronym{MCP}{MCP}{Model Context Protocol}
\newacronym{OWL}{OWL}{Web Ontology Language}
\newacronym{MES}{MES}{Manufacturing Execution System}
\newacronym{SCADA}{SCADA}{Supervisory Control and Data Acquisition}
\newacronym{RPM}{RPM}{revolutions per minute}

\begin{document}
\bstctlcite{IEEEexample:BSTcontrol} 

\title{Beyond Formal Semantics for Capabilities and Skills: Model Context Protocol in Manufacturing}

\author{
\IEEEauthorblockN{
    Luis Miguel Vieira da Silva\IEEEauthorrefmark{1},
    Aljosha Köcher\IEEEauthorrefmark{1},
    Felix Gehlhoff\IEEEauthorrefmark{1}
}
\IEEEauthorblockA{
\IEEEauthorrefmark{1}Institute of Automation Technology\\
Helmut Schmidt University, Hamburg, Germany\\
Email: \{miguel.vieira, aljosha.koecher, felix.gehlhoff\}@hsu-hh.de\\}
}

\maketitle

\begin{abstract}
Explicit modeling of capabilities and skills -- whether based on ontologies, Asset Administration Shells, or other technologies -- requires considerable manual effort and often results in representations that are not easily accessible to \glspl{LLM}. 
In this work-in-progress paper, we present an alternative approach based on the recently introduced \gls{MCP}. 
\gls{MCP} allows systems to expose functionality through a standardized interface that is directly consumable by \gls{LLM}-based agents. 
We conduct a prototypical evaluation on a laboratory-scale manufacturing system, where resource functions are made available via \gls{MCP}. 
A general-purpose \gls{LLM} is then tasked with planning and executing a multi-step process, including constraint handling and the invocation of resource functions via \gls{MCP}. The results indicate that such an approach can enable flexible industrial automation without relying on explicit semantic models. 
This work lays the basis for further exploration of external tool integration in \gls{LLM}-driven production systems.
\end{abstract}


\section{Introduction}
\label{sec:introduction}
\glsresetall
Modern industrial automation systems are increasingly expected to operate in dynamic environments, where production requirements, tasks, and system configurations change frequently. 
To meet these demands, such systems must be both modular and reconfigurable \cite{ElM_Flexibleandreconfigurablemanufacturing_2005}. 

A common approach to describe and manage the functional potential of these systems is the use of \glspl{CSS}. 
Capabilities describe abstract, implementation-independent functions, whereas skills represent their concrete implementations and are exposed via a skill interface \cite{KBH+_AReferenceModelfor_15.09.2022b}. 
This separation supports abstraction, reuse, and planning by allowing reasoning on the capability level while delegating execution to the skill level.
However, the practical use of capabilities and skills in flexible automation faces a persistent challenge in the formal modeling of capabilities and skills. This model creation is labor-intensive, error-prone, and demands deep semantic modeling expertise \cite{LAF+_TowardaMethodto_2024}. 

At the same time, \glspl{LLM} have emerged as powerful reasoning tools capable of analyzing complex problems, generating plans, and supporting decision-making processes. 
However, \glspl{LLM} do not natively have access to functionalities like machine interfaces, control systems, or sensor values. 
To integrate external functionalities with \glspl{LLM}, \gls{MCP} was recently introduced \cite{Ant_ModelContextProtocolSpecification_2025}. 
\glspl{LLM} can interpret task descriptions, reason about available functionalities, select and invoke suitable functions via \gls{MCP}, and thus contribute to flexible planning and control.

This paper investigates how capabilities and skills can be represented, exposed, and utilized in an \gls{LLM}-\gls{MCP}-based automation architecture. 
Specifically, we address the following research questions:
\begin{itemize}
\item \textbf{RQ1:} How can \gls{MCP} be used to make capabilities and skills available for \glspl{LLM} so that their reasoning powers can be used for manufacturing planning and control?
\item \textbf{RQ2:} Can formal models of \gls{CSS} be replaced by natural language descriptions that are interpretable by \glspl{LLM}?
\end{itemize}

The following Section~\ref{sec:fundamentals} gives an introduction to \gls{MCP}, while Section~\ref{sec:relatedWork} reviews related work on \gls{MCP} and use of \glspl{LLM} in industrial automation. 
Section~\ref{sec:approach} introduces our approach for representing and deploying capabilities and skills in an \gls{MCP}-compatible format. 
Section~\ref{sec:evaluation} presents an experimental setup and preliminary evaluation results. 
Finally, Section~\ref{sec:conclusion} concludes the paper with an outlook on future research directions.

\section{Model Context Protocol} 
\label{sec:fundamentals}
\glsreset{MCP}
\textbf{\gls{MCP}}, introduced in \cite{Ant_ModelContextProtocolSpecification_2025}, is an open standard for connecting \glspl{LLM} to external resources, prompt templates and tools, i.e., executable functions, in a structured and unified way. 
While an \gls{LLM} alone may conceptually plan a task for a robotic cell, it cannot efficiently execute the corresponding low-level operations without a standardized mechanism to invoke such external functions.
With \gls{MCP}, a protocol layer is provided that allows external functions to be described, discovered, and invoked at runtime -- without requiring prior hard-coded integration into the \gls{LLM}.

\gls{MCP} follows a client-server architecture: \gls{MCP} servers expose functional interfaces, metadata, and invocation mechanisms; \gls{MCP} clients, e.g., \gls{LLM}-based agents, can dynamically query these servers to discover available operations and execute them by issuing well-defined requests. 
\cite{Ant_ModelContextProtocolSpecification_2025}

One of \gls{MCP}’s key strengths is the decoupling of models and external functions: \glspl{LLM} do not need to be trained with specific tool knowledge. 
Instead, they can query the available \gls{MCP} servers during execution and decide based on context and goals which functionality to use. This makes \gls{MCP} highly suitable for dynamic, modular environments.

\section{Related Work} 
\label{sec:relatedWork}
A recent survey in \cite{Ray_ASurveyonModel_2025} outlines \gls{MCP}’s core concepts and its potential to expose external tools to \glspl{LLM} in a standardized way. 
The authors discuss possible applications of \gls{MCP} in domains such as manufacturing, where \gls{LLM} agents might retrieve sensor data, interact with \gls{MES} platforms, or initiate control actions in response to anomalies. 
However, Ref.~\cite{Ray_ASurveyonModel_2025} highlights only conceptual opportunities and does not present concrete implementations or evaluate the feasibility of using \glspl{LLM} for tool selection or task execution in physical automation systems.

While the authors of \cite{Ray_ASurveyonModel_2025} outline the conceptual potential of the new \gls{MCP} for exposing tools to \glspl{LLM}, other recent studies focus more directly on how \glspl{LLM} can be applied to planning and control tasks in industrial automation.
Xia et al. present in \cite{XJZ+_ControlIndustrialAutomationSystem_2024} an agent-based framework using structured prompts, event-driven semantics, and a hierarchy of manager and operator agents that interact with automation modules via digital twins. 
The control relies on generated function calls and standard operation procedures as a knowledge base, but requires fine-tuning, curated datasets, and tightly engineered prompt structures -- limiting flexibility and generalizability.

A similar direction is pursued in the LLMAPM framework introduced in \cite{NWL+_Alargelanguagemodelbased_2025}, which targets manufacturing process planning. 
Natural language task descriptions are decomposed into workflows, verified through state machines, and executed on heterogeneous devices via a low-code platform. While supporting human-in-the-loop corrections, the system still depends on handcrafted prompt templates and lacks a runtime interface for tool abstraction and reuse.

In \cite{VyMe_AutonomousIndustrialControlusing_2024} a multi-agent \gls{LLM} control framework uses validator and reprompter agents with a digital twin to iteratively refine actions for robust decision-making in industrial settings. 
While this enables safe and adaptive control, it is demonstrated on a single closed-loop task and does not generalize to dynamic function composition or modular tool invocation.

The use of \glspl{LLM} in multi-robot systems is reviewed by Li et al. in \cite{LZA+_LargeLanguageModelsfor_2025}, covering applications in task allocation and coordination. 
While relevant for distributed systems, the focus remains on agent negotiation and motion planning rather than exploring how \glspl{LLM} might interface with standardized protocols like \gls{MCP} for structured tool invocation.

A broader perspective on \gls{LLM}-driven agent systems is provided in \cite{LiTo_LLMPoweredAIAgentSystems_2025}, which categorizes \gls{LLM}-powered agents into software-based, physical, and adaptive hybrid types. 
While the survey highlights potential use cases in smart manufacturing, it acknowledges ongoing challenges such as output uncertainty, integration complexity, and lack of standardized tool access in physical environments.

In contrast to these works, our approach operationalizes \gls{MCP} in an industrial context by exposing machine-level functions as callable tools and describing them in natural language. 
This enables \glspl{LLM} to perform planning and control tasks without fine-tuning, handcrafted prompts, or formal semantic models -- offering a scalable and lightweight alternative for flexible automation.

\section{MCP Approach for Flexible Manufacturing}
\label{sec:approach}
\begin{figure*}[htb]
    \centering
    \includegraphics[width=0.7\textwidth]{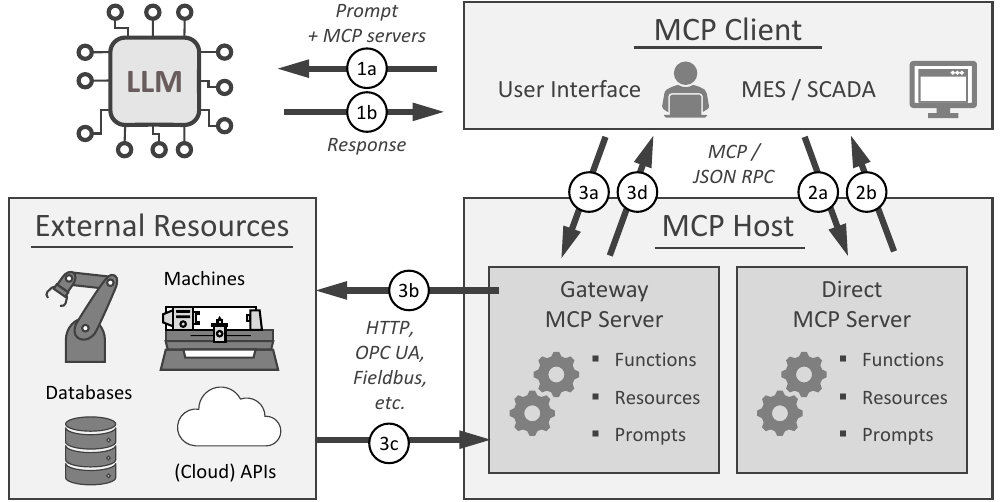}
    \caption{Overview of an MCP architecture for manufacturing.}
    \label{fig:approachOverview}
    \vspace{-0.4em}
\end{figure*}

The goal of this paper is to present a first examination for a lightweight alternative to formal and explicit models of capabilities and skills. Current approaches use ontologies, \glspl{AAS}, or other technologies, which always require a significant effort for creating and maintaining information models. 
The core idea of this paper is to leverage \gls{MCP} to expose and describe capabilities and skills in a way that can easily be consumed and reasoned about by \glspl{LLM}. 
This enables \glspl{LLM} not only to understand available functionalities, but also to orchestrate and execute them without the need for predefined information models.

The intended objectives of this approach directly mirror the objectives of conventional modeling approaches for capabilities and skills:
\begin{enumerate}
    \item \textbf{Capability Discovery and Representation}: Operators and control systems, such as \gls{MES}, require an overview of available capabilities and their constraints, as well as access to executable skills. \gls{MCP} serves as a standardized protocol to expose this information dynamically.
    \item \textbf{Automatic Skill Invocation}: With given interface descriptions provided via \gls{MCP}, an \gls{LLM} can evaluate the structure of a skill function, supply suitable parameters, trigger its execution, and react on returned results. 
    This replaces rigid model-based invocation logic with a more flexible, language-driven interaction and \gls{LLM} reasoning.
    \item \textbf{Service-Level Negotiation}: \glspl{LLM} can act as intelligent service requesters that formulate services with required capability descriptions and dynamically compare the service offers given by multiple service providers without the need for predefined negotiation mechanisms.
\end{enumerate}

\subsection{MCP Architecture}
In the proposed architecture (see Figure~\ref{fig:approachOverview}), user interfaces and control systems like \gls{MES} act as \gls{MCP} clients that are directly connected to an \gls{LLM}.
\glspl{MCP} servers provide functions, resources, and prompts. 
According to the definition in~\cite{KBH+_AReferenceModelfor_15.09.2022b}, a capability is defined as a specification of a function with either physical or virtual effects.
This distinction of effects is reflected in the \gls{MCP} architecture with the two alternatives of direct \gls{MCP} servers and gateway \gls{MCP} servers.
\begin{itemize}
    \item \textbf{Direct \gls{MCP} Servers} represent purely virtual or computational capabilities with skills implemented and executed locally on the \gls{MCP} host. They may include calculations, conversions, data analyses, or decision logic.
    \item \textbf{Gateway \gls{MCP} Servers} serve as intermediary functions to trigger or interact with data and functionalities on external resources. This includes physical capabilities, i.e., real world manufacturing functions as well as requesting information from external APIs or databases.
    To invoke skills on external resources, gateway \gls{MCP} servers act as a wrapper that invokes actual control logic via interfaces such as HTTP, OPC UA, or ROS~2.
\end{itemize}
User queries or control requests are sent together with all available \gls{MCP} servers as prompts to the \gls{LLM} (1a).
Depending on the response of the \gls{LLM} (1b), \gls{MCP} functionalities are selected and used by the \gls{MCP} client. In case of direct \gls{MCP} servers, this functionality is called (2a) and the returned result (2b) is interpreted. 
In the case of gateway \gls{MCP} servers, the function call (3a) may lead to an external request (3b) that yields a return (3c), which is processed by the gateway \gls{MCP} server and then returned to the \gls{MCP} client (3d).
The following subsection illustrates how elements of the \gls{CSS} reference model presented in \cite{KBH+_AReferenceModelfor_15.09.2022b} relate to elements in this \gls{MCP} architecture.

\subsection{Representing CSS Elements in MCP}
Each \textbf{Capability} is a function declaration in one of the programming frameworks provided by \cite{Ant_ModelContextProtocolSpecification_2025}. Its specification details are documented in natural language using descriptions and code comments. This natural-language description is the counterpart of a capability model and enables \glspl{LLM} to understand the purpose, constraints, and semantics of a capability.

\textbf{Properties} of capabilities are represented by the arguments and return values of the capability function. The data types, units, and purposes of properties are expressed as code comments. A property may directly map to a SkillParameter, though some arguments may serve purely descriptive purposes (e.g., to express constraints that are not used in execution).

\textbf{Constraints} are subdivided into PropertyConstraints and TransitionConstraints \cite{KBH+_AReferenceModelfor_15.09.2022b}. PropertyConstraints are internal to a capability. They can be expressed as comments and embedded as assertions or preconditions inside the capability function. Violations can raise structured errors, which the LLM can interpret (e.g., drill diameter $\leq 50$).
TransitionConstraints are global constraints between multiple capabilities: These cannot be checked by one capability alone, but instead must be described textually (e.g., ``may only be used after Capability X'') and are interpreted by the \gls{LLM} during invocation.

A \textbf{Skill} is the implementation of a capability, i.e., the function definition. This is in line with the specification of a skill in \cite{KBH+_AReferenceModelfor_15.09.2022b}. However, \gls{MCP} provides an alternative interface to expose and describe the skill’s callable entry points.

\textbf{State Machines} define the internal logic, state transitions, and safety behavior of a skill and remain unchanged compared to a skill implementation with an explicit information model. \gls{MCP} does not modify the underlying control logic.

The explicit modeling of a \textbf{SkillInterface} is replaced by code-level documentation and the generic \gls{MCP} interaction logic based on JSON-RPC. Instead of explicitly modeling each method or parameter, the LLM infers available operations from code comments and \gls{MCP} metadata.

\textbf{SkillParameters} are all variables used in the function implementation. In certain cases, all capability properties, i.e., function arguments, are used as skill parameters. 
In other scenarios, additional variables (e.g., external configuration values) may further parameterize skill behavior without being capability properties.

\section{Evaluation}
\label{sec:evaluation}
To validate our approach, we conducted a proof-of-concept implementation using real automation components. 
The goal was to assess whether an \gls{LLM} can autonomously reason over a task description, select and orchestrate appropriate capabilities via \gls{MCP} tools, and execute the skill process. 
All experiments were performed using \emph{Claude Desktop} as the \gls{MCP} client with the \gls{LLM} \emph{Sonnet 4} and \glspl{MCP} servers hosted on the same computer. The results are available online\footnote{\url{https://github.com/CaSkade-Automation/css-mcp-study}}.
We implemented three \gls{MCP} tools corresponding to three capabilities:
\begin{itemize}
    \item \textbf{Spindle Speed Calculation}: A direct \gls{MCP} server with a Python script that selects a recommended spindle speed for drilling based on material type and drill diameter from a predefined lookup table.
    \item \textbf{Drilling}: A gateway \gls{MCP} server to trigger a drilling operation on a real production system via OPC UA.
    \item \textbf{Mobile Robot Transport}: A ROS~2-based capability interfacing with a Neobotix MMO-700 mobile robot to pick up and transport workpieces between stations.
\end{itemize}
All capabilities were exposed as \gls{MCP} tools with code comments as described in Section~\ref{sec:approach}. 
We designed four evaluation scenarios that showcase manufacturing tasks requiring planning, conditional logic, and interpretation of results:

\begin{enumerate}
    \item \textbf{Standard Process Execution}: A user specifies a workpiece to be drilled (i.e. defines material and diameter) and transported to an assembly station. The \gls{LLM} correctly infers that the spindle speed must be calculated first. It uses the calculation skill to retrieve a \gls{RPM} value, invokes the drilling skill with this \gls{RPM} value, and dispatches the mobile robot for transport. All skills were executed in correct order and with correct parameters.
    \item \textbf{Unsupported Drill Diameter}: A user provides an invalid diameter not listed in the \gls{RPM} table. The \gls{LLM} correctly interprets the error response and reacts by prompting the user to select a supported drill diameter. After correction, the process continues as expected.
    \item \textbf{Workpiece at Different Location}: The scenario is identical to the first one, but the workpiece is initially at a different station. The \gls{LLM} infers the need for a transport step before drilling. It first uses the robot to transport the part, then proceeds like in Scenario 1.
    \item \textbf{Material Name Variation}: A user specifies a material not included in the lookup table ("stainless steel" instead of "stainless"). The calculation skill returns an error due to unmatched material input. The \gls{LLM} identifies "stainless" as a likely alternative and successfully retries the request with the corrected term.
\end{enumerate}

Across all scenarios, the \gls{LLM} successfully planned and executed multi-step workflows by selecting and invoking the appropriate \gls{MCP} tools with correct parameters at runtime. 
The system demonstrated robust handling of ambiguous or invalid inputs, including skill result interpretation, dynamic re-querying, and user clarification. 
Notably, no hard-coded control flow was required -- the capability orchestration was a result of the \gls{LLM}’s reasoning based on capability descriptions and real-time interaction with skills.

\section{Conclusion and Future Work}
\label{sec:conclusion}
This paper presents a lightweight and \gls{LLM}-driven alternative to current model-based \gls{CSS} approaches in manufacturing. 
By leveraging \gls{MCP}, we demonstrate how capabilities can be described using comments and skills can be exposed as callable tools (RQ1).
Through a set of scenarios, we demonstrate the potential of an \gls{LLM} paired with \gls{MCP} to plan and execute multi-step manufacturing processes, discover available capabilities, interpret their purposes, invoke corresponding skills, and handle errors or ambiguities gracefully -- without explicit information models, e.g., ontologies or \glspl{AAS}. 
Our results suggest that explicit \gls{CSS} models may become less necessary in the future, as \glspl{LLM} increasingly infer intent from text and function implementations (RQ2).
This would significantly lower the barrier to integration and reconfiguration in flexible manufacturing.

However, our results are only an initial validation of \gls{LLM}-based orchestration via \gls{MCP}, and reliability and robustness -- including hallucinations -- remain an open challenge with many directions for future work.
Instead of a single \gls{LLM} orchestrator, future approaches should use multiple, dedicated \gls{LLM}-based agents with defined roles (e.g., planner, verifier, negotiator) for more scalable, modular, and robust architectures.
In addition, decentralized negotiation patterns to compare service offers across organizational boundaries may be realized by \gls{LLM} agents.

Another future research direction is to enable \glspl{LLM} not only to select and invoke skills, but to also generate new \emph{ad-hoc skills} on demand. Building on earlier work on skill generation \cite{LAN+_CapabilityDriven_2025}, a combined approach could allow \glspl{LLM} to identify missing functionality, synthesize \gls{MCP} tool code, register the new tool, and invoke it as part of a task plan. The approach of "model-free" automation, where planning and execution emerge from linguistic reasoning rather than predefined symbolic models, can also be applied to simplify integration processes in other domains, e.g., modular electrolyzer systems.

\section*{Acknowledgment}
This research is funded by dtec.bw – Digitalization and Technology Research Center of the Bundeswehr as part of the project RIVA. dtec.bw is funded by the European Union – NextGenerationEU. The authors also gratefully acknowledge the funding of the project eModule (Support code: 03HY116) by the German Federal Ministry of Education and Research, based on a resolution of the German Bundestag 

\bibliographystyle{./bibliography/IEEEtran}
\bibliography{./bibliography/references} 

\end{document}